\documentstyle[11pt,newpasp,epsf,twoside]{article}
\def\mh{{M_{\bullet}}}
\def\mb{{M_{\rm bulge}}}
\def\msun{{M_{\odot}}}
\def\mpc{{\rm Mpc}}
\def\ms{{\mh-\sigma}}
\def\rh{{r_h}}
\def\etal{{\it et al.\ }}

\def\ml{{\mh-M_B}}

\def\Sec{\hbox{${}^{\prime\prime}$\llap{.}}}

\def\sc{\hbox{${}^{\hbox{\sevenrm s}}$\llap{.}}}

\def\kms{km s$^{-1}$}

\def\fun#1#2{\lower3.6pt\vbox{\baselineskip0pt\lineskip.9pt
  \ialign{$\mathsurround=0pt#1\hfil##\hfil$\crcr#2\crcr\sim\crcr}}}
\def\lap{\mathrel{\mathpalette\fun <}}
\def\gap{\mathrel{\mathpalette\fun >}}

\def\edcomment#1{\iffalse\marginpar{\raggedright\sl#1\/}\else\relax\fi}
\reversemarginpar

\begin{document}
\title{Relationship of Black Holes to Bulges}
 \author{David Merritt and Laura Ferrarese}
\affil{Department of Physics and Astronomy, Rutgers University, New Brunswick, NJ, USA}

\begin{abstract}
Supermassive black holes appear to be uniquely associated with
galactic bulges.
The mean ratio of black hole mass to bulge mass was until recently
very uncertain, with ground-based, stellar kinematical data
giving a value for $\langle\mh/\mb\rangle$ roughly an order
of magnitude larger than other techniques.
The discrepancy was resolved with the discovery of the $\ms$ relation,
which simultaneously established a tight corrrelation between black
hole mass and bulge velocity dispersion, and confirmed that the
stellar kinematical mass estimates were systematically too large due to
failure to resolve the black hole's sphere of influence.
There is now excellent agreement between the various techniques 
for estimating $\langle\mh/\mb\rangle$,
including dynamical mass estimation in quiescent galaxies; 
reverberation mapping in active galaxies and quasars;
and computation of the mean density of compact objects based on 
integrated quasar light.
All techniques now give $\langle\mh/\mb\rangle\approx 10^{-3}$ and
$\rho_{\bullet}\approx 3\times 10^5\msun/{\rm Mpc}^{-3}$.
Implications of the $\ms$ relation for the formation of black holes
are discussed.
\end{abstract}

\section{Introduction}

The argument that active galaxies and quasars are powered
by accretion onto massive compact objects was first made
almost four decades ago (Salpeter 1964; Zeldovich 1964).
Since that time, the existence of supermassive black holes 
has been confirmed in the nuclei of
nearby galaxies and in a handful of more distant galaxies
by direct dynamical measurement of their masses.
The best determinations are in our own Galaxy
($\mh\approx 3\times 10^6\msun$, Genzel \etal 2000;
Ghez \etal 1998), NGC 4258 (Miyoshi \etal 1995),
and M87 (Macchetto \etal 1997).
The data for each of these galaxies exhibit a clear rise 
very near the center in the orbital velocity of stars or gas,
suggestive of motion around a compact object.
Data of this quality are unfortunately still rare, 
and the majority of black hole detections have necessarily 
been based on stellar-kinematical data which do not
exhibit a clear signature of the presence of a black hole.
These data (Magorrian \etal 1998; Richstone \etal 1998) 
imply a mean black hole mass 
that is uncomfortably large compared with values predicted from 
quasar light.
The inconsistency has been taken as evidence for low radiative
efficiencies during the optically bright phase of quasars
(e.g. Haehnelt, Natarajan \& Rees 1998)
or for continued growth of black holes after the quasar epoch
(e.g. Richstone \etal 1998).

It is now clear that this discrepancy was due almost entirely
to systematic errors in the stellar kinematical mass estimates.
The first convincing demonstration of this came from
the $\ms$ relation,
a tight empirical correlation between black hole
mass and bulge velocity dispersion.
The $\ms$ relation was discovered
by ranking black hole detections in terms of their believability
and excluding the least secure cases.
The remarkable and unexpected result (Ferrarese \& Merritt 2000) 
was an almost perfect correlation between $\mh$ and $\sigma$ 
for the best-determined black hole masses, compared
to a much weaker correlation for the less secure masses.
Ground-based, stellar kinematical estimates of $\mh$
were found to scatter above the $\ms$ relation by as much as 
two orders of magnitude, suggesting that many of the published 
masses were spurious and that most were substantial overestimates.

The ability of the $\ms$ relation to ``separate the wheat from
the chaff'' has led to a rapid advance in our understanding of 
black hole demographics.
We review that progress in \S2 and \S3; we argue that there is now
almost embarrassingly good agreement between the results from the various
techniques for estimating the mean mass density of black holes
in the universe.
Black hole masses determined dynamically in nearby quiescent 
galaxies are now fully consistent with masses inferred 
in active galaxies and quasars,
and with estimates of the density of dark relic objects
produced by accretion during the quasar epoch.
The need for non-standard accretion histories in order to reproduce 
a large density of black holes in the current universe has disappeared.

Two recurrent themes in this review are the importance of
adequate data when estimating black hole masses;
and the much greater usefulness of {\it accurate} mass estimates
compared with simple detections.
When the first black hole detections were being published,
there was much discussion about whether the observations 
(all ground-based at the time) 
were of sufficient quality to resolve the black holes' 
sphere of influence, $\rh=G\mh/\sigma^2$.
We now know that the ground-based data almost always failed
to do this, sometimes by a large factor, and that this
failure, coupled with shortcomings in the modelling,
led to systematic overestimates of $\mh$ (\S2).
The situation has improved somewhat with the Space Telescope, 
but not dramatically:
we argue (\S 4) that the number of galaxies with secure dynamical
estimates of $\mh$ will increase only modestly over the next few
years in spite of ambitious ongoing programs with HST.
This is due partly to the fact that these observations were
planned at a time when the mean black hole mass was believed
to be much larger than it is now.
Progress in our understanding of black hole demographics is 
more likely to come from techniques with higher effective resolution
than stellar or gas kinematics,
notably reverberation mapping in AGN (Peterson 1993).

While the ability of the $\ms$ relation to clarify the data has
been an enormous boon,
the existence of such a tight correlation must also be telling us something 
fundamental about the way in which black holes form and about the
connection between black holes and bulges.
Unfortunately, the theoretical understanding of this connection has
lagged behind the phenomenology.
We summarize the proposed explanations for the origin of the
$\ms$ relation in \S 5 and discuss their strengths and weaknesses.

\section{Black Hole Mass Estimates: A Critical Review}

\begin{quote}
{\it Epimetheus}: Wie vieles ist denn dein?\\
\noindent{\it Prometheus}: Der Kreis, den meine Wirksamkeit erf\"ullt!

\noindent {\it Epimetheus}: What then do you possess?\\
\noindent {\it Prometheus}: My sphere of influence - nothing more and nothing less!

{\it Goethe, Prometheus}
\end{quote}

\subsection{A Discrepancy, and its Resolution}

By 1999, a clear discrepancy was emerging between black hole
masses derived from stellar kinematical studies and 
most other techniques.
The former sample included many ``standard bearers'' 
like M31 (Richstone, Bower \& Dressler 1990),
NGC 3115 (Kormendy \etal 1996a) and NGC 4594 (Kormendy \etal 1996b).
The size of the discrepancy was difficult to pin down since
there were (and still are) essentially no galaxies for which black 
hole masses had been independently derived using more than one technique.
However the masses derived from ground-based stellar kinematics
were much larger, by roughly an order of magnitude on average, 
than those inferred from other techniques when galaxies with similar properties
were compared, or when estimates of the cosmological density of
black holes or the mean ratio of black hole mass to bulge mass were made.
The discrepancy was clearest in two arenas:

\begin{itemize}

\item {\it Active vs. Quiescent Galaxies.}
In Seyfert galaxies, 
the gravitational influence of the black hole
can be measured on small scales, $\sim 10^{-2}$ pc,
using emission lines from the broad emission line region (BLR).
The technique of ``reverberation mapping'' combines the velocity
of the BLR gas with an estimate of the size of the BLR 
based on time delay measurements (Peterson 1993).
Reverberation mapping masses have been derived for about 
three dozen black holes (Wandel, Peterson \& Malkan 1999;
Kaspi \etal 2000).
These masses fall a factor of 5-20 below the $\mh-L_{\rm bulge}$
relation defined by ground-based, stellar kinematical data 
(Wandel 1999, 2000).

\item{\it Quasar Light vs. Black Hole Demographics.}
The mass density of black holes at large redshifts
can be estimated by requiring the optical QSO luminosity 
function to be reproduced by accretion onto black holes
(Soltan 1982).
Assuming a standard accretion efficiency of $\sim 10\%$,
the mean mass density in black holes works out to be
$\rho_{\bullet}\sim 2\times 10^5\msun\mpc^{-3}$
(Chokshi \& Turner 1992; Salucci \etal 1999).
A similar argument based on the X-ray background gives
consistent results,
$\rho_{\bullet}\sim 3\times 10^5\msun \mpc^{-3}$ 
(Fabian \& Iwasawa 1999; Salucci \etal 1999).
By comparison, the black hole mass density implied by 
stellar kinematical modelling was about ten times higher
(Magorrian \etal 1998; Richstone \etal 1998; Faber 1999).

\end{itemize}

Serious inconsistencies like these only appeared when comparisons were
made with black hole masses derived from the stellar kinematical data; 
all other techniques gave roughly consistent values for 
$\rho_{\bullet}$ and $\langle\mh/M_{\rm bulge}\rangle$.
Nevertheless, most authors accepted the correctness of the
stellar dynamical mass estimates and looked elsewhere to explain 
the discrepancies.
Ho (1999) suggested that the reverberation mapping masses had been
systematically underestimated.
Wandel (1999) proposed that black holes in active galaxies were
smaller on average than those in quiescent galaxies, due either
to different accretion histories or to selection effects.
Richstone \etal (1998) and Faber (1999) suggested that the inconsistency
between their group's masses and the masses inferred from 
quasar light 
could be explained if black holes had acquired $80\%$ of 
their mass {\it after} the quasar epoch
through some process that produced no observable radiation.

\begin{figure}
\plotfiddle{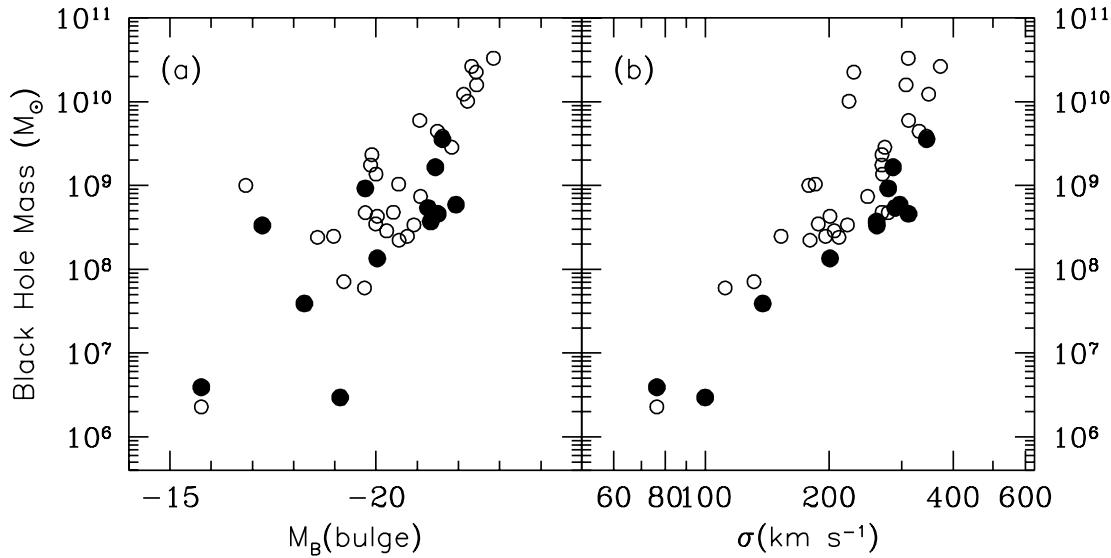}{2.75truein}{0.}{75.}{75.}{-230}{-120}
\caption{Discovery of the $\ms$ relation 
(adapted from Ferrarese \& Merritt 2000).
(a) $\mh$ vs. bulge luminosity; (b) $\mh$ vs. bulge velocity dispersion.
``Sample A'' masses, derived from high quality data, are indicated
with filled circles; 
``Sample B'' masses, from lower quality data, are the open circles.
}
\end{figure}

What particularly caught our attention was the gulf between black hole
masses derived from high- and low-resolution data, and
(to a lesser extent) between gas- and stellar dynamical data;
the former (e.g. Ferrarese, Ford \& Jaffe 1996; Ferrarese \& Ford 1999)
were typically taken at higher resolution than the latter.
Black hole masses derived from the highest resolution data, 
in galaxies like the Milky Way (Ghez \etal 1998; Genzel \etal 2000)
and M87 (Macchetto \etal 1997), were the 
{\it smallest} when expressed as a fraction of the bulge mass, 
with $\mh/\mb\approx 10^{-3}$.
The {\it largest} fractional black hole masses -- in galaxies like
NGC 3377 (Kormendy \etal 1998) or NGC 4486b (Kormendy \etal 1997) -- 
were mostly derived from stellar absorption-line spectra obtained from the 
ground, at resolutions of $\sim1''$, corresponding to typical
linear scales of $10-100$ pc.
The mean value of $\mh/\mb$ for these galaxies was claimed
to be about $10^{-2}$ (Magorrian \etal 1998; Richstone \etal 1998), 
roughly an order of magnitude
greater than the value derived from the high-resolution data.
We began to suspect that some of the masses derived from the
lower-quality data might be serious over-estimates
-- or, even worse, that some of the ``detections'' based on these 
data were spurious.

To test this hypothesis, we tabulated all of the published black
hole masses that had been derived from stellar- or gas kinematical
data (excluding the reverberation mapping masses) and divided
them into two groups based on their expected accuracy.
This is not quite as easy as it sounds, since the ``accuracy'' 
of a black hole mass estimate is not necessarily related 
in any simple way to its published confidence range.
Our criterion was simply the quality of the data:
``accurate'' black hole masses were those derived
from HST data, at resolutions of $\sim 0.1''$,
as well as $\mh$ for the Milky Way black hole (which is by far the 
nearest) and the black hole in NGC 4258 (for which VLBI gives
a resolution of $\sim 0.1$ pc).
The velocity data for these galaxies (our ``Sample A'') was always 
found to exhibit a clear rise in the inner few data points,
suggesting that the black hole's sphere of influence
$\rh\equiv G\mh/\sigma^2$ had been well resolved.
The remaining black hole masses (``Sample B'') were all those 
derived from  lower-resolution data, typically ground-based
stellar kinematics, including most of the masses in the
Magorrian \etal (1998) study.
Sample A contained 12 galaxies, Sample B 31.

Our first attempt to compare ``Sample A'' and ``Sample B'' masses
was disappointing (Figure 1a).
In the $\mh-L_{\rm bulge}$ plane, 
the Sample A masses do fall slightly below those from Sample B,
but the intrinsic scatter in $L_{\rm bulge}$ is apparently so large 
that there is no clear difference in the relations defined by the
two samples.

But when we plotted $\mh$ versus the velocity dispersion $\sigma$
of the bulge stars, something magical happened (Figure 1b): now the
Sample A galaxies clearly defined the {\it lower edge} of the 
relation, while the Sample B galaxies scattered above, 
some by as much as two orders of magnitude in $\mh$!
Furthermore the correlation defined by the Sample A galaxies 
alone was very tight.

What particularly impressed us about the $\mh-\sigma$ plot was the fact that
the Sample A galaxies, which are diverse in their properties,
showed such a tight correlation; while the Sample B galaxies, which are much
more homogeneous, exhibited a large scatter.
For instance, Sample A contains two spiral galaxies, two lenticulars,
and both dwarf and giant ellipticals; while the Sample B
galaxies are almost exclusively giant ellipticals.
Furthermore the black hole masses in Sample A were derived using
a variety of techniques, including absorption-line stellar kinematics 
(M32, NGC 4342), 
dynamics of gas disks (M87, NGC 4261), 
and velocities of discrete objects (MW, NGC 4258);
while in Sample B all of the black hole masses were derived from
stellar spectra obtained from the ground.
This was circumstantial, but to us compelling, evidence
that the Sample A masses were defining the true relation and
that the Sample B masses were systematically in error.

Fitting a regression line to $\log\mh$ vs. $\log\sigma$ for
the Sample A galaxies alone, we found
\begin{equation}
\mh = 1.40 \times 10^8 \msun \left({\sigma_c\over 200\ {\rm km}\
{\rm s^{-1}}}\right)^{\alpha}
\end{equation}
with $\alpha=4.80\pm 0.5$ (Ferrarese \& Merritt 2000).
We defined the quantity $\sigma_c$ to be the rms velocity of stars 
in an aperture
of radius $r_e/8$ centered on the nucleus, with $r_e$ the half-light
radius of the bulge.
This radius is large enough that the stellar
velocities are expected to be affected at only the few percent
level by the gravitational force from the black hole,
but small enough that $\sigma_c$ can easily be measured
from the ground.

A striking feature of the $\ms$ relation is its negligible
scatter.
The reduced $\chi^2$ of Sample A about the best-fit line of Eq. 1, 
taking into account measurement errors in both variables, 
is only $0.74$, essentially a perfect fit.
Such a tight correlation seemed almost too good to be true
(and may in fact be a fluke resulting from the small sample size)
but we felt we could not rule it out given the existence of
other, similarly tight correlations in astronomy, e.g. 
the near-zero thickness of the elliptical galaxy fundamental plane.

\begin{figure}
\plotone{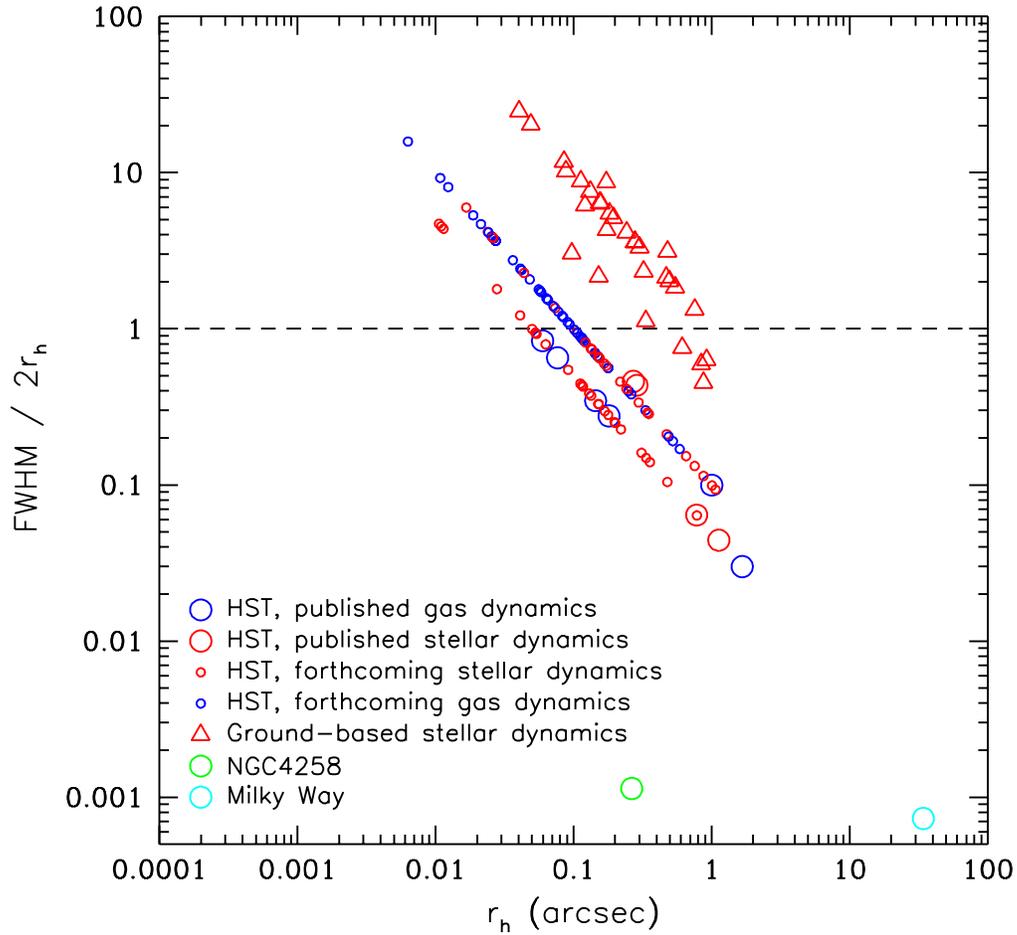}
\caption{Detectability of black holes in galaxies for which dynamical
mass determinations have been published or are planned.
Horizontal axis is $\rh\equiv G\mh/\sigma^2$, the black hole's
radius of influence. $\mh$ is computed from the $\ms$ relation
for all forthcoming and ground-based observations.
Vertical axis is the ratio of the observational resolution to
twice $\rh$.
Determination of $\mh$ is difficult when this ratio is $\gap 1$,
and mass determinations based on stellar dynamics (red symbols) can
be difficult even when ${\rm FWHM}/2\rh<1$, for reasons discussed
in the text.
}
\end{figure}

In fact the scatter in the $\ms$ relation is so small that it is
reasonable to use the relation to {\it predict} black hole masses, 
even in galaxies for which determinations of $\mh$
based on detailed modelling have previously been published.
One can then ask, galaxy by galaxy, 
whether the observations on which the published
estimate of $\mh$ was based were of sufficiently high quality to 
resolve the black hole's sphere of influence.
Table 1 and Figure 2 show the results.
Table 1 is a ranked list of the most secure black hole detections
to date.
The galaxies are listed in order of increasing
${\rm FWHM}/2r_h$, i.e. the ratio of the size of the resolution
element to twice the radius of influence of the black hole. 
In the case of HST observations, for which the PSF is undersampled, FWHM
is the diameter of the FOS aperture or the width of the STIS slit.
For ground-based observations,  FWHM refers to the seeing
disk. 
Figure 2 plots the same quantities for essentially all galaxies with 
published estimates of $\mh$ based on stellar or gas kinematics.

Not surprisingly, only the black holes in the Milky Way 
and in NGC 4258 have been observed at a resolution
greatly exceeding $\rh$.
The Sample A galaxies of Ferrarese \& Merritt (2000)
also satisfy ${\rm FWHM}/2\rh<1$, although sometimes marginally.
By contrast, almost none of the ground-based data resolved
$\rh$, sometimes failing by more than a factor of 10.

\scriptsize
\begin{table}
\begin{tabular}{l l r l r l l}
\multicolumn{7}{c}{\bf \normalsize Ranked Census of Supermassive Black Hole Detections \tablenotemark{1,2}}\\
\multicolumn{7}{c}{}\\
\hline\hline
\multicolumn{1}{c}{Galaxy} &
\multicolumn{1}{c}{Type} &
\multicolumn{1}{c}{Distance} &
\multicolumn{1}{c}{$\mh$} &
\multicolumn{1}{c}{$\sigma_c$} &
\multicolumn{1}{c}{${\rm FWHM}/2\rh$} &
\multicolumn{1}{c}{Reference}\\
\hline
\multicolumn{7}{c}{}\\
\multicolumn{7}{c}{\it \normalsize Galaxies for which $\rh$ has been resolved}\\
\hline
MW    & SbI-II      & 0.008  & 0.0295$\pm$0.0035        & 100$\pm$20 & $7.3^{-4}$ & Genzel \etal 2000\\ 
N4258 & SAB(s)bc    & 7.2    & 0.390$\pm$0.034          & 138$\pm$18 & $1.1^{-3}$ & Miyoshi \etal 1995\\   
N4486 & E0pec       & 16.7   & 35.7$\pm 10.2$           & 345$\pm$45 & 0.03 & Macchetto \etal 1997\\   
N3115 & S0$^-$      & 9.8    & 9.2$\pm$3.0              & 278$\pm$36 & 0.04 & Emsellem \etal 1999\\
N221  & cE2         & 0.8    & 0.039$\pm$0.009          & 76$\pm$10  & 0.06 & Joseph \etal 2000\\  
N5128 & S0pec       & 4.2    & 2.4$^{3.6}_{-1.7}$       & 145$\pm$25 & 0.10 & Marconi \etal 2001\\
N4374 & E1          & 18.7   & 17$^{+12}_{-6.7}$        & 286$\pm$37 & 0.10 & Bower \etal 1998\\   
N4697 & E6          & 11.9   & 1.7$^{+0.2}_{-0.3}$      & 163$\pm$21 & 0.10 & ``Nuker'' group, unpubl.\tablenotemark{3}\\
N4649 & E2          & 17.3   & 20.6$^{+5.2}_{-10.2}$    & 331$\pm$43 & 0.10 & ``Nuker'' group, unpubl.\tablenotemark{3}\\
N4261 & E2          & 33.0   & 5.4$^{+1.2}_{-1.2}$      & 290$\pm$38 & 0.18 & Ferrarese \etal 1996\\   
M81   & SA(s)ab     & 3.9    & 0.68$^{0.07}_{-0.13}$    & 174$\pm$17 & 0.19 & STIS IDT, unpubl.\tablenotemark{3} \\  
N4564 & E           & 14.9   & 0.57$^{+0.13}_{-0.17}$   & 153$\pm$20 & 0.33 & ``Nuker'' group, unpubl.\tablenotemark{3}\\
I1459 & E3          & 30.3   & 4.6$\pm$2.8              & 312$\pm$41 & 0.35 & Verdoes Kleijn \etal 2000\\   
N5845 & E*          & 28.5   & 2.9$^{+1.7}_{-2.7}$      & 275$\pm$36 & 0.40 & ``Nuker'' group, unpubl.\tablenotemark{3}\\
N3379 & E1          & 10.8   & 1.35$\pm$0.73            & 201$\pm$26 & 0.44 & Gebhardt \etal 2000a\\
N3245 & SB(s)b      & 20.9   & 2.1$\pm$0.5              & 211$\pm$19 & 0.48 & Barth \etal 2001\\
N4342 & S0$^-$      & 16.7   & 3.3$^{+1.9}_{-1.1}$      & 261$\pm$34 & 0.56 & Cretton \& van den Bosch 1999\\   
N7052 & E           & 66.1   & 3.7$^{+2.6}_{-1.5}$      & 261$\pm$34 & 0.66 & van der Marel \& van den Bosch 1998\\   
N4473 & E5          & 16.1   & 0.8$^{+1.0}_{-0.4}$      & 188$\pm$25 & 0.77 & ``Nuker'' group, unpubl.\tablenotemark{3}\\
N6251 & E           & 104    & 5.9$\pm$2.0              & 297$\pm$39 & 0.84 & Ferrarese \& Ford 1999\\ 
N2787 & SB(r)0+     & 7.5    & 0.41$^{0.04}_{-0.05}$    & 210$\pm$23 & 0.87 & Sarzi \etal 2001\\
N3608 & E2          & 23.6   & 1.1$^{+1.4}_{-0.3}$      & 206$\pm$27 & 0.98 & ``Nuker'' group, unpubl.\tablenotemark{3}\\  
\hline
\multicolumn{7}{c}{}\\
\multicolumn{7}{c}{\it \normalsize Galaxies for which $\rh$ has not been resolved}\\
\hline
N3384 & SB(s)0$^-$  & 11.9   & 0.14$^{+0.05}_{-0.04}$   & 151$\pm$20  & 1.0 & ``Nuker'' group, unpubl.\tablenotemark{3}\\
N4742 & E4          & 15.5   & 0.14$^{0.04}_{-0.05}$    &  93$\pm$10  & 1.0 & STIS IDT, unpubl.\tablenotemark{3} \\
N1023 & S0          &10.7    & 0.44$\pm0.06         $   & 201$\pm$14  & 1.1 & STIS IDT, unpubl.\tablenotemark{3}\\
N4291 & E           & 26.9   & 1.9$^{+1.3}_{-1.1}$      & 269$\pm$35  & 1.1 & ``Nuker'' group, unpubl.\tablenotemark{3}\\
N7457 & SA(rs)0$^-$ & 13.5   & 0.036$^{+0.009}_{-0.011}$& 73$\pm$10   & 1.1 & ``Nuker'' group, unpubl.\tablenotemark{3}\\ 
N821  & E6          & 24.7   & 0.39$^{+0.17}_{-0.15}$   & 196$\pm$26  & 1.3 & ``Nuker'' group, unpubl.\tablenotemark{3} \\
N3377 & E5+         & 11.6   & 1.10$^{+1.4}_{-0.5}$     & 131$\pm$17  & 1.3 & ``Nuker'' group, unpubl.\tablenotemark{3}\\
N2778 & E           & 23.3   & 0.13$^{+0.16}_{-0.08}$   & 171$\pm$22  & 2.8 & ``Nuker'' group, unpubl.\tablenotemark{3}\\
\hline
\multicolumn{7}{c}{}\\
\multicolumn{7}{c}{\it \normalsize Galaxies in which dynamical studies are inconclusive}\\
\hline
N224  & \multicolumn{5}{l}{Double nucleus, system not in dynamical equilibrium.}& Bacon \etal 2001 \\
N598 & \multicolumn{5}{l}{Data imply upper limit only, $\mh\lap10^3\msun$.}& Merritt, Ferrarese \& Joseph 2001\\
N1068 & \multicolumn{5}{l}{Velocity curve is sub-Keplerian.}& Greenhill \etal 1996\\
N3079 & \multicolumn{5}{l}{Masers do not trace a clear rotation curve.} & Trotter et al. 1998\\
N4459 & \multicolumn{5}{l}{Data do not allow unconstrained fits.}& Sarzi \etal 2001\\
N4486B& \multicolumn{5}{l}{Double nucleus, system not in dynamical equilibrium.}& STIS IDT, unpubl.\tablenotemark{2}\\
N4945 & \multicolumn{5}{l}{Asymmetric velocity curve; velocity is sub-Keplerian.}& Greenhill \etal 1997\\
\hline\hline
\end{tabular}
\tablenotetext{1}{Type is revised Hubble type.  Black hole
masses are in $10^8$ solar masses, velocity dispersions are in km
s$^{-1}$, and distances are in Mpc.  $\sigma_c$ is the
aperture-corrected velocity dispersion defined by Ferrarese \& Merritt
(2000).  $\rh=G\mh/\sigma_c^2$, with $\mh$ the value in column 4.
References in column 7 are to the papers in which the dynamical
analysis leading to the mass estimate were published.  }
\tablenotetext{2}{For the reasons outlined in the text,the masses from
Magorrian et al. (1998) are omitted from this tabulation. This
includes NGC 4594, which was included in Kormendy \& Gebhardt (2001).}
\tablenotetext{3}{Preliminary masses tabulated in Kormendy \& Gebhardt
(2001). Data and modelling for these mass estimates are not yet
available.}
\end{table}
\normalsize

The latter point is important, since precisely these
data were used to define the canonical relation between
black hole mass and bulge luminosity
(Magorrian \etal 1998; Richstone \etal 1998; Faber 1999)
that has served as the basis for so many subsequent studies
(e.g. Haehnelt, Natarajan \& Rees 1998; Catteneo, Haehnelt \& Rees 1999; 
Salucci \etal 1999; Kauffmann \& Haehnelt 2000; 
Merrifield, Forbes \& Terlevich 2000).
Figure 3 plots the likely ``error'' in the ground-based mass estimates
(defined as the ratio of the quoted mass, 
$M_{\rm fit}$, to the mass implied by Eq. 1)
as a function of the effective resolution ${\rm FWHM}/2\rh$.
The error 
is found to correlate strongly with the quality of the data.
For the best-resolved of the Magorrian \etal candidates,
${\rm FWHM}/2\rh\lap 1$,
the average error in $\mh$ appears to be a factor of $\sim 3$, 
rising roughly linearly with ${\rm FWHM}/\rh$ to values of
$\sim 10^2$ for the most poorly-resolved candidates.

\begin{figure}
\plotone{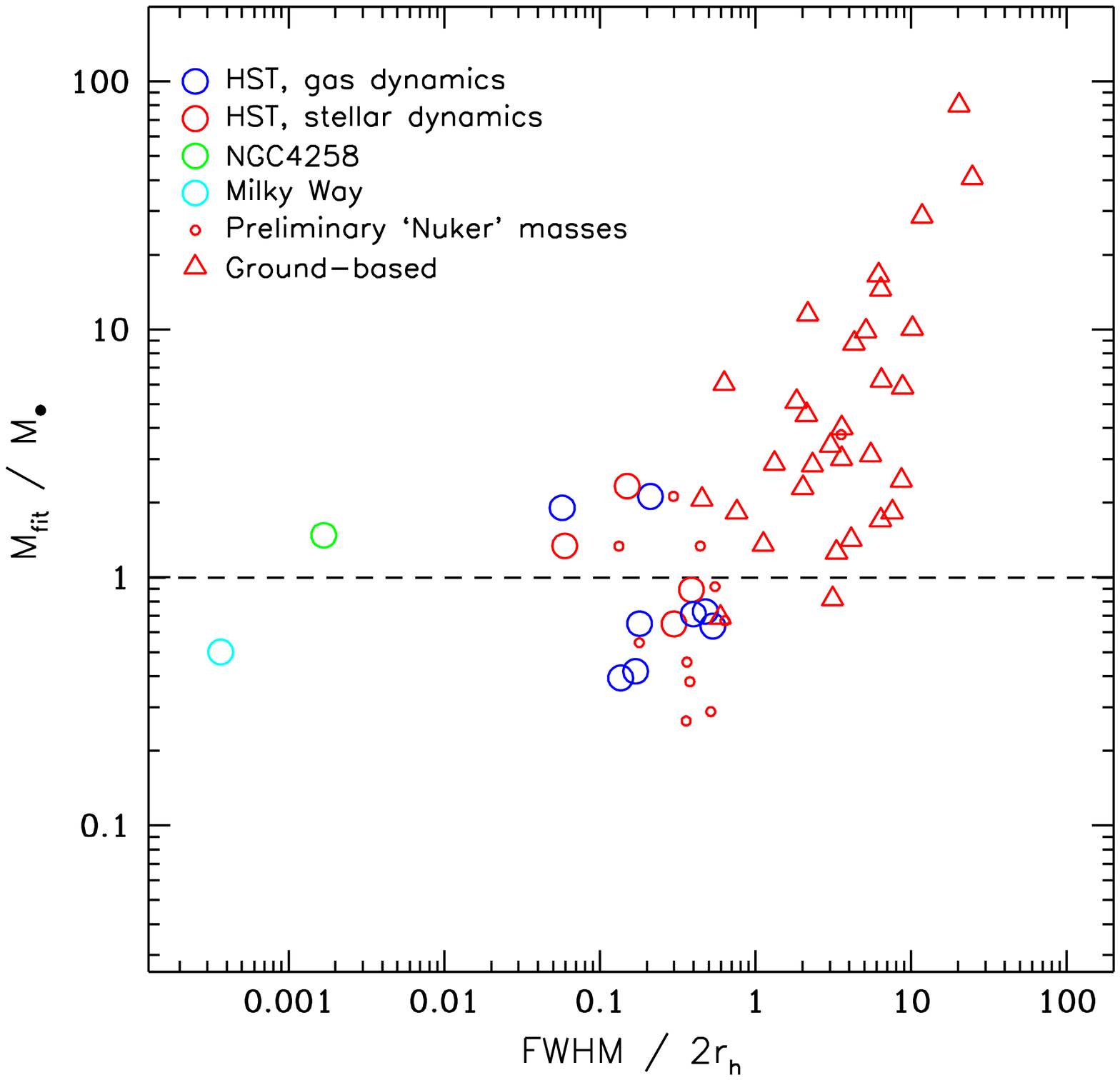}
\caption{
``Error'' in published black hole masses, defined
as the ratio of the published mass estimate $M_{\rm fit}$
to the value $\mh$ inferred from the $\ms$ relation, as a function of the
effective resolution of the data from which the mass estimate was derived.
The error increases roughly inversely with resolution for
${\rm FWHM}/2\rh\gap 1$.
Most of the ground-based detections are in this regime.
}
\end{figure}

An important quantity is the mean ratio of black hole
mass to bulge mass, $\langle\mh/\mb\rangle$.
Figure 4 compares the distribution of $M_{\rm fit}/\mb$,
the mass ratio computed by Magorrian \etal (1998),
to the distribution obtained when $M_{\rm fit}$ is replaced 
by $\mh$ as computed from the $\ms$ relation.
The mean value of $(\mh/\mb)$ drops from $1.7\times 10^{-2}$
to $2.5\times 10^{-3}$, roughly an order of magnitude.
The mean value of $\log_{10}(\mh/\mb)$ shifts downward by $-0.7$
corresponding to a factor $\sim 5$ in $\mh/\mb$.
The density of black holes in the local universe implied by the
lower value of $\langle\mh/\mb\rangle$ is
$\rho_{\bullet}\sim 5\times 10^5\msun {\rm Mpc}^{-3}$ 
(Merritt \& Ferrarese 2001a), consistent with the value
required to explain quasar luminosities assuming
a standard accretion efficiency of 10\% (Chokshi \& Turner 1992;
Salucci \etal 1999; Barger \etal 2001).
 
\begin{figure}
\plotfiddle{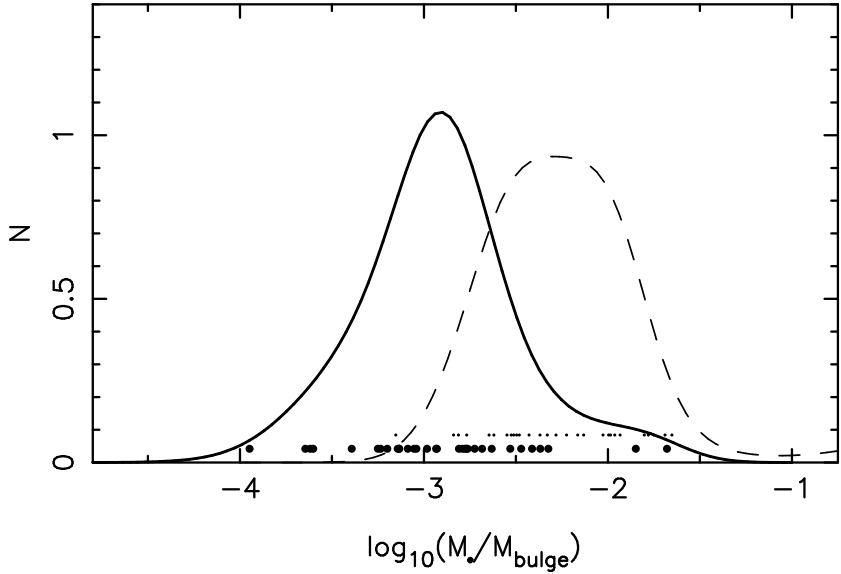}{2.7truein}{0.}{60.}{60.}{-190}{-100}
\caption{
Frequency function of black-hole-to-bulge mass ratios
(adapted from Merritt \& Ferrarese 2001a).
The dashed curve is the ``Magorrian relation'' (Magorrian \etal 1998)
based on black hole masses derived from ground-based kinematics
and two-integral modelling.
The solid curve is the frequency function obtained when 
black hole masses are instead computed from the $\ms$ relation.
}
\end{figure}

\subsection{Pitfalls of Stellar Dynamical Mass Estimation}

Why were most of the stellar dynamical mass estimates
so poor; why were they almost always over-estimates;
and what lessons do past mistakes have for the future?
The answer to the first question is simple in retrospect.
Figure 5 shows how the signal of the black hole -- a
sudden rise in the rms stellar velocities at a distance of
$\sim \rh\equiv G\mh/\sigma^2$ from the black hole -- 
is degraded by seeing.
For ${\rm FWHM}/2\rh\gap 2$, the signal is so small as to be 
almost unrecoverable except with data of exceedingly high
$S/N$.
Most of the ground-based observations fall into this regime
(Figure 2).
In fact the situation is even worse than Figure 5 suggests, 
since for ${\rm FWHM}\gap \rh$, the rise in $\sigma(R)$ will
be measured by only a single data point.
This is the case for many of the galaxies that are listed
as ``resolved'' in Table 1 (e.g. NGC 3379, Gebhardt \etal 2000a).

\begin{figure}
\plotfiddle{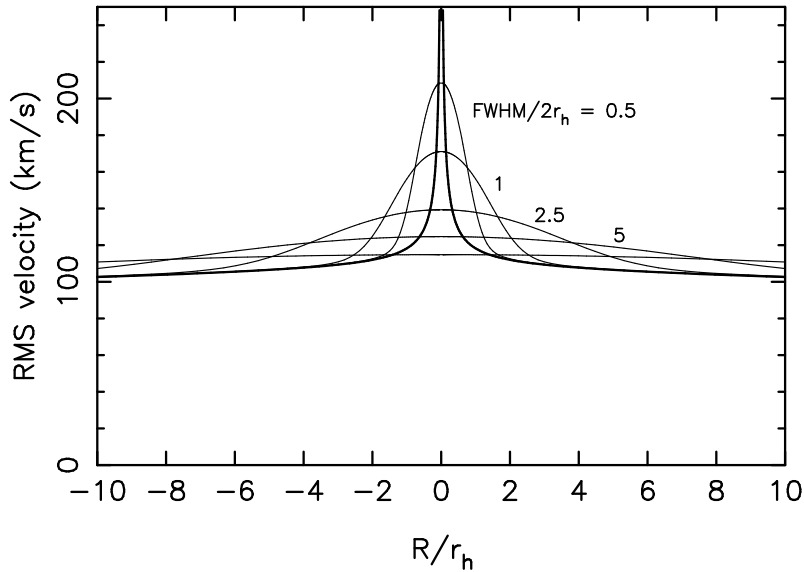}{2.7truein}{0.}{60.}{60.}{-205}{-240}
\caption{
Degradation due to seeing of the velocity dispersion spike 
produced by a black hole in a hot (nonrotating) stellar system.
Heavy line is the profile unaffected by seeing; $R$ is the
projected distance from the black hole and $\rh=G\mh/\sigma^2$.
When ${\rm FWHM}/2\rh\gap\sim 2$, the velocity dispersion
spike is so degraded as to be almost unrecoverable.
Most ground-based observations fall into this regime
(Fig. 2).}
\end{figure}

A short digression is in order at this point. 
Data taken from the ground often show an impressive central
spike in the velocity dispersion profile; examples are
NGC 4594 (Kormendy \etal 1996b) and  NGC 4486b (Kormendy \etal 1997).
However such features are due in part to blending of light from two 
sides of the nucleus where the rotational velocity has opposite signs
and would be almost as impressive even if the black hole 
were not present.
This point was first emphasized by Tonry (1984) in the context of
his ground-based M32 observations.
As he showed, the velocity dispersion spike in M32 as observed
at $\sim 1''$ resolution is consistent with rotational broadening 
and does not require any increase in the {\it intrinsic} velocity dispersion 
near the center.

Why should poor data lead preferentially to overestimates of $\mh$,
rather than random errors?
There are two reasons.
First, as pointed out by van der Marel (1997), much of the
model-fitting prior to 1999 was carried out
using isotropic spherical models or their axisymmetric analogs,
the so-called ``two-integral'' (2I) models.
Such models predict a velocity dispersion profile
that gently {\it falls} as one moves inward, 
for two reasons: non-isothermal cores, i.e. $\rho\sim r^{-\gamma}$
with $\gamma\ne\{0,2\}$, generically have central minima in the rms velocity
(e.g. Dehnen 1993); and, when flattened, the 2I axisymmetric models
become dominated by nearly circular orbits (in order to maintain
isotropy in the meridional plane) further reducing the predicted
velocities near the center. 
Figure 6 illustrates these effects for a set of axisymmetric 2I models with
$\gamma=1.5$.
Real galaxies almost always exhibit a monotonic rise in $v_{\rm rms}$.
Adding a central point mass can correct this deficiency of the models,
but only an unphysically large value of $M_{\rm fit}$ will affect the 
stellar motions at large enough radii, $r\lap 0.1r_e$, 
to do the trick.
This is probably the explanation for the factor $\sim 3$ mean error 
in $\mh$ derived from the best ground-based data (Figure 3).

\begin{figure}
\plotfiddle{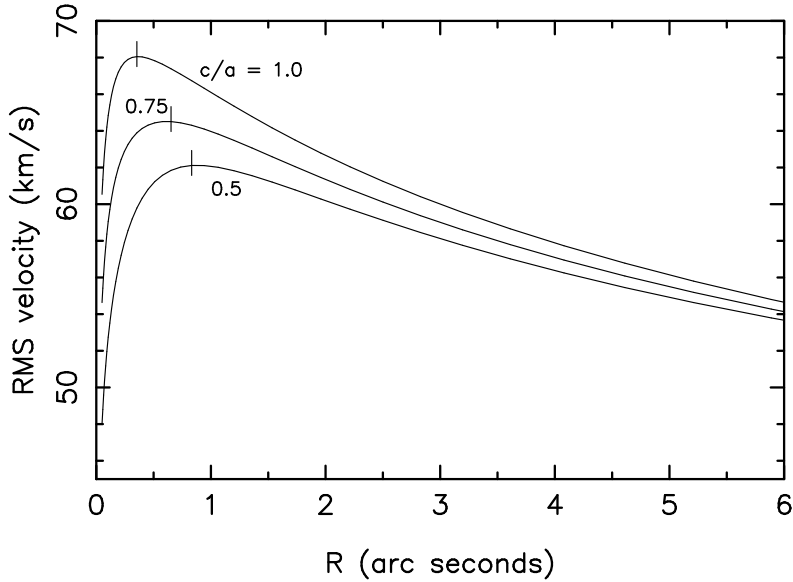}{2.7truein}{0.}{60.}{60.}{-205}{-240}
\caption{Velocity dispersion profiles of the ``two-integral''
(2I) models that were used as templates for estimating black hole
masses in many of the stellar kinematical studies (e.g. Magorrian
\etal 1998).
Model flattening is indicated as $c/a$; there are no central black holes.
Ticks mark the point $R_{max}$ of maximum velocity;
$R_{max}$ moves outward as the flattening is increased.
}
\end{figure}

The much larger values of $M_{\rm fit}/\mh$ associated with
the more distant galaxies in Figure 3 are probably attributable to 
a different factor.
When the data contain no useful information about the black hole mass,
only values of $M_{\rm fit}$ that are much larger than the true 
mass will significantly affect the $\chi^2$ of the model fits.
The only black holes that can be ``seen'' in such data are 
excessively massive ones.

Can these problems be overcome by abandoning 2I models in favor
of more general, three-integral (3I) models?
The answer, surprisingly, is ``no'': making the modelling algorithm
more flexible (without also increasing the amount or quality of the
data) has the effect of {\it weakening} the constraints 
on $\mh$.
The reason is illustrated in Figure 7.
The rms velocities in 2I models are uniquely
determined by the assumed potential, i.e. by $M_{\rm fit}$ and
$M/L$, the mass-to-light ratio assumed for the stars.
This means that the models are highly {\it over-constrained} by the
data -- there are far more observational constraints (velocities)
than adjustable parameters ($M_{\rm fit},M/L$), hence one expects to find
a unique set of values for $M_{\rm fit}$ and $M/L$ that come closest to
reproducing the data.
This is the usual case in problems of statistical estimation
and it implies a well-behaved set of $\chi^2$ contours with
a unique minimum.

When the same data are modeled using the more general distribution 
of orbits available in a 3I model, 
the problem becomes {\it under}-constrained:
now one has the freedom to adjust the phase-space distribution
function in order to compensate for changes in $M_{\rm fit}$ and $M/L$, 
so as to leave the goodness of fit precisely unchanged.
The result is a plateau in $\chi^2$ (Figure 7), the width
of which depends in a complicated way on the ratio of 
observational constraints to number of orbits or phase-space cells
in the modelling algorithm (Merritt 1994).
Thus, 3I modelling of the ground-based data would
only show that the range of possible values of $M_{\rm fit}$ 
includes, but is not limited to, the values found using the 2I models;
it would not generate more precise estimates of $\mh$ unless
the data quality were also increased.

\begin{figure}
\plotfiddle{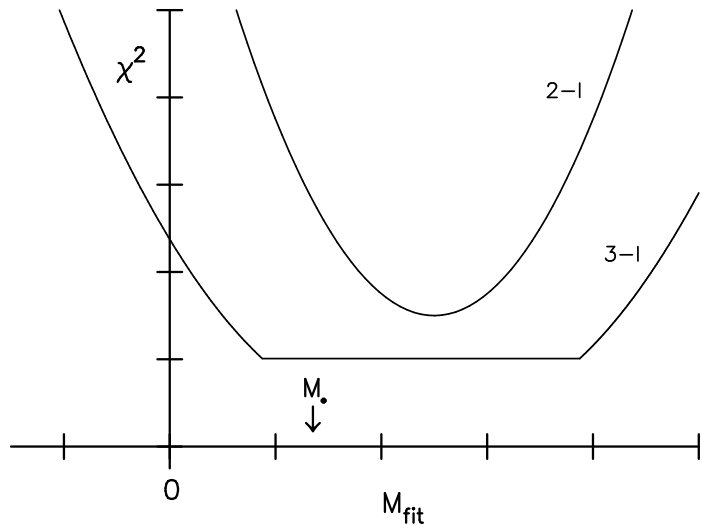}{2.6truein}{0.}{60.}{60.}{-205}{-210}
\caption{Schematic comparison of two-integral (2I) and
three-integral (3I) modelling of stellar kinematical data;
$M_{\rm fit}$ is the estimated black hole mass. 
2I models predict a unique rms velocity field given an assumed mass
distribution; 
in 3I models, the extra freedom associated with a more
general distribution of orbits allows one to compensate for
changes in $M_{\rm fit}$ in such a way as to leave the goodness of fit to the
data precisely unchanged.
}
\end{figure}

The greater difficulty of interpreting results from 3I modelling 
has not been widely appreciated; few authors make
a distinction between ``indeterminacy'' in $\mh$
(the width of the constant-$\chi^2$ plateau in Figure 7)
and ``uncertainty'' (the additional range in  
$\mh$ allowed by measurement errors), or look carefully
at how their confidence range depends on the number
of orbits used.
We illustrate these difficulties by examining two recently published
studies based on high quality, stellar kinematical data.

1. {\sl NGC 3379} (Gebhardt \etal 2000a): The prima-facie evidence
for a central mass concentration in this galaxy consists of a single
data point, the innermost velocity dispersion as measured by HST/FOS;
the rotation curve exhibits no central rise,
in fact it drops monotonically toward the center.
Goodness-of-fit contours generated from 3I models show the expected 
plateau (Fig. 7 of Gebhardt \etal),
extending from $\sim 10^6\msun$ to $\sim 3\times 10^8\msun$.
In fact a model with $\mh=0$ fits the data just as well: the
authors state that ``the difference between the no-black hole and
black hole models is so subtle'' as to be almost indiscernable
(cf. their Fig. 11).
Gebhardt \etal nevertheless argue for $\mh>0$ based on the poorly-determined
wings of stellar velocity distribution measured within
the central FOS resolution element.
In view of the fact that this velocity distribution
exhibits a puzzling unexplained asymmetry (their Fig. 4), 
the stellar dynamical case for a black hole
in this galaxy should probably be considered marginal.

2. {\sl NGC 4342} (Cretton and van den Bosch 1999): The evidence
for a central mass concentration is again limited to a single data
point, the central FOS velocity dispersion.
Cretton \& van den Bosch find that a black-hole-free model provides
``fits to the actual data [that] look almost indistinguishable
from that of Model B'' (a model with $M_{\rm fit}=3.6\times 10^8\msun$).
Their $\chi^2$ contours (their Fig. 7) nevertheless seem to show a preferred
black hole mass; however they note that $\chi^2$ is dominated by the data
at radii $R\gap 5''$, far outside of the radius of influence of the
black hole.
The probable culprit here is the modest number of orbits ($1400$,
compared with $\sim 250$ constraints) in their 3I solutions.
Outer data points are always the most difficult to fit when
modelling via a finite orbit library since only a fraction of the orbits
extend to large radii; this is clear in their fits (cf. their Fig. 8) 
which become progessively worse at large radii.

We emphasize that both of these modelling studies were based on
high-quality data, with ${\rm FWHM}/2\rh\approx 0.4$ (NGC 3379)
and $0.6$ (NGC 4342) (Table 1).
Nevertheless, the extreme freedom associated with 3I models permits a 
wide range of black hole masses to be fit to the
velocity data in both galaxies.
As Figure 2 shows, most of the galaxies in the ongoing HST/STIS survey
of galactic nuclei will be observed at even lower effective resolutions;
hence we predict that the black hole masses in many of these galaxies
will turn out to be consistent with zero and that the range of allowed
masses will usually be large.
(To be fair, we note that these observations were planned at a time when 
$\langle\mh/\mb\rangle$ was believed to be much larger than it is now.)
We therefore urge caution when interpreting results like 
Kormendy \& Gebhardt's (2001) recent compilation
of black hole masses derived from unpublished 3I modelling.


\section{Supermassive Black Holes in Active Galactic Nuclei}

The techniques that allow us to detect supermassive black holes
in quiescent galaxies are rarely applicable to the hosts of bright AGNs.  
In the Seyfert 1 galaxies and in the handful of QSOs that are close 
enough that the black hole's sphere of influence has some chance
of being resolved, the presence of the bright non-thermal
nucleus (e.g. Malkan, Gorjian \& Tam 1998) severely dilutes the very
features which are necessary for dynamical studies.  The only
bright AGN in which a supermassive black hole has been detected by 
spatially-resolved kinematics is the nearby (Herrnstein et al. 1999;
Newman \etal 2000) Seyfert 2 galaxy NGC 4258, which is
blessed with the presence of an orderly water maser disk (Watson \& Wallin 1994;
Greenhill \etal  1995; Miyoshi \etal  1995).  The radius of
influence of the black hole at its center, $\sim$0\Sec15, can barely be
resolved by HST but can be fully sampled by the VLBA at 22.2 GHz.  
Unfortunately, water masers are rare and of the handful that are known,
only in NGC 4258 are the maser clouds distributed in a simple geometrical 
configuration that exhibits clear Keplerian motion around the central source 
(Braatz \etal 1996; Greenhill \etal 1996, 1997;
Greenhill, Moran \& Herrnstein 1997; Trotter \etal 1998).  
Black hole demographics in AGNs must therefore proceed via
alternate routes.

Dynamical modeling of the broad emission line region (BLR) constitutes
a viable alternative to spatially-resolved kinematical studies. 
According to the standard model, the BLR consists of many ($10^{7-8}$,
Arav \etal  1997, 1998; Dietrich \etal  1999), small, dense ($N_e
\sim 10^{9-11}$ cm$^{-3}$), cold ($T_e \sim 2\times10^4$ K)
photoionized clouds (Ferland \etal  1992), localized within a volume
of a few light days to several tens of light weeks in diameter around the
central ionization source (but see also Smith \& Raine 1985,
1988; Pelletier \& Pudritz 1992; Murray \etal  1995; Murray \& Chiang
1997; Collin-Souffrin \etal  1988).  As such, the BLR is, and will
likely remain, spatially unresolved.  In the presence of a
variable non-thermal nuclear continuum, however, the
responsivity-weighted radius $R_{BLR}$ of the BLR is measured by the
light-travel time delay between emission and continuum variations
(Blandford \& McKee 1982; Peterson 1993; Netzer \& Peterson 1997;
Koratkar \& Gaskell 1991).  {\it If} the BLR is gravitationally bound,
the central mass is given by the virial theorem as $M_{\rm virial} = v_{BLR}^2
R_{BLR} / G$, where the FWHM of the emission lines (generally
H$\beta$) is taken as being representative of the rms velocity
$v_{BLR}$, once assumptions are made about the BLR geometry.  In a few
cases, independent measurements of $R_{BLR}$ and $v_{BLR}$ have been
derived from different emission lines: it is found that the two
quantities define a ``virial relation'' in the sense 
$v_{BLR}\sim r^{-1/2}$ (Koratkar \& Gaskell 1991;
Wandel, Peterson \& Malkan 1999; Peterson \& Wandel 2000), suggesting
a simple picture of a stratified BLR in Keplerian motion.  

On the downside, mapping the BLR response to continuum variations
requires many ($\sim 10^{1-2}$) repeated observations taken at closely
spaced time intervals, $\Delta t \lap 0.1 R_{BLR}/c$.
Moreover, the observations can be translated into black hole
masses only if a series of reasonable, but untested, 
assumptions are made regarding the geometry, stability and
velocity structure of the BLR, 
the radial emissivity function of the gas, and the
geometry and location (relative to the BLR) of the ionizing continuum
source.  If a wrong assumption is made, systematic errors of 
a factor $\sim 3$ can result (Krolik 2001).
The uncertainties surrounding reverberation mapping 
 has made the derived black hole masses an easy target for critics 
(e.g. Richstone \etal 1998; Ho \etal  1999).
On the other hand, because the BLR gas samples a spatial region very
near to the black hole, there is almost no possibility of making
the much larger errors in $\mh$ that have plagued the
ground-based stellar kinematical studies (Magorrian \etal 1998).
Thanks to the efforts of international collaborations, 
reverberation mapping masses are now available for 17
Seyfert 1 galaxies and 19 QSOs (Wandel, Peterson \& Malkan 1999; Kaspi
\etal 2000).

Taken at face value, reverberation mapping radii are found to
correlate with  the non-thermal optical luminosity of the nuclear
source. While the exact functional form of the dependence is debated  
(Koratkar \& Gaskell 1991; Kaspi \etal  1996, 2000;
Wandel, Peterson \& Malkan 1999), the $R_{BLR} - L$ relation can
potentially provide an inexpensive way of bypassing reverberation
mapping measurements on the way to determining black hole masses.

\subsection{AGN Black Hole Demographics from the $\mh-\mb$ Relation}

With one exception (Ferrarese \etal 2001), black hole demographic 
studies for AGNs have been based on the $\ml$, rather than on the 
$\ms$, relation for the simple reason that few accurate $\sigma$ measurements 
exist in AGN hosts (e.g. Nelson \& Whittle 1995). $L_{\rm bulge}$, on the other hand, 
is more easily measured than $\sigma$ (though not
necessarily more {\it accurately} measured, as discussed below).
The modest sample of AGNs with reverberation mapping black hole masses is
often augmented using masses derived from the $R_{BLR} - L$
relation (Wandel 1999; Laor1998, 2000; McLure \& Dunlop 2000).
For a sample of 14 PG quasars, Laor (1998) reported reasonable
agreement with the $\ml$ relation derived by Magorrian \etal (1998)
for quiescent galaxies, finding $\langle\mh/\mb\rangle=0.006$.   
Seyfert 1 galaxies define a significantly different
correlation according to Wandel (1999): $\langle \mh/\mb\rangle=0.0003$.  
Most recently, McLure \& Dunlop (2000) have reanalyzed the QSO sample of Laor 
and the Seyfert sample of Wandel (the first augmented with almost as many new
objects and both  with new spectroscopic and/or photometric data for
the existing objects). McLure \& Dunlop split the difference of the two
ealier studies by obtaining $\langle\mh/\mb\rangle=0.0025$.
They find no statistical difference between
Seyfert 1s and QSOs. 

The different conclusions reached by these authors can be traced to
a number of factors.

\begin{itemize}

\item Bulge magnitudes are at the heart of the problem for the Seyfert
sample (McLure \& Dunlop 2000; Laor 2001).  Wandel used luminosities
derived (by Whittle \etal  1992) using the Simien \& de Vaucouleurs
(1986) empirical correlation between galaxy type and bulge/disk ratio.
HST images allowed McLure \& Dunlop to perform a proper disk/bulge
decomposition, which produces bulges $1-3.5$ magnitudes fainter than
assumed by Wandel (1999), hence larger $\mh/\mb$.

\item Overestimated black hole masses seem to be responsible for the large
$\langle\mh/\mb\rangle$ measured by Laor. 
Here, the bolometric non-thermal nuclear
luminosity used in  estimating $R_{BLR}$ from the $R_{BLR}-L$ relation
is a factor $\sim  3$ larger in Laor than in McLure \& Dunlop (for the
same cosmology).  Everything else being equal, this leads to a factor
$\sim 2$ increase  in the black hole masses. It is unclear which
luminosities are more correct; 
however it seems that the McLure \& Dunlop values are
to be preferrred  for the following reason. The $R_{BLR}-L$ relation
is defined using  monochromatic luminosity (at 5100 \AA~in Kaspi et
al. 2000, and 4800  \AA~in Kaspi \etal 1996). This can be transformed
to a bolometric  luminosity by assuming a power law of given spectral
index for the  nuclear spectrum. McLure \& Dunlop used monochromatic
luminosities  applied to the Kaspi \etal (2000) relation, while Laor
started from  bolometric luminosities (from Neugebauer \etal 1987),
applied a  constant bolometric correction, and then used the Kaspi et
al. (1996)  relation. The more direct route used by McLure \& Dunlop
(which  bypasses the need for a bolometric correction) seems to be
preferable.

\end{itemize}

\begin{figure}
\plotone{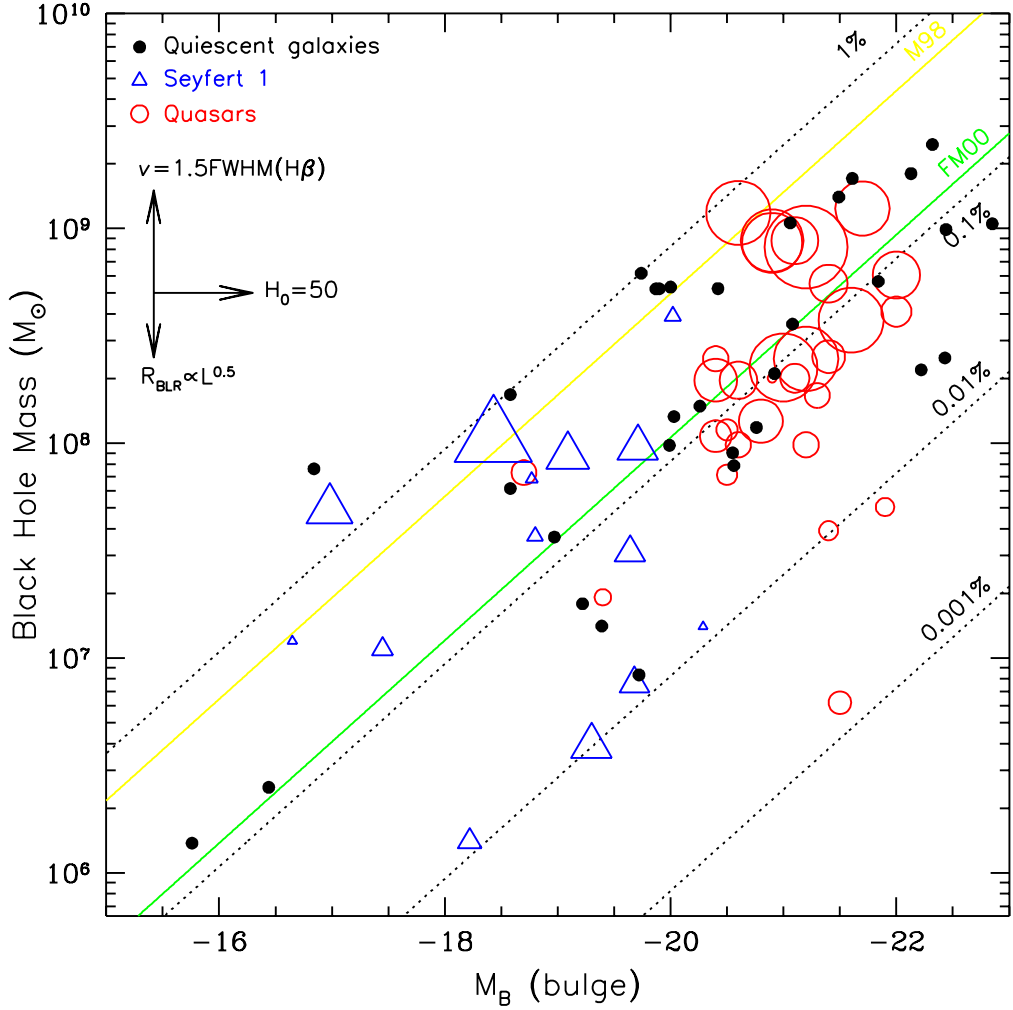}
\caption{The $\mh/\mb$ relation for quiescent galaxies
(solid black dots, with best fit given by the green line), Seyfert 1
galaxies (blue triangles) and nearby QSOs (red circles). The data are
taken from Ferrarese \& Merritt (2000), McLure \& Dunlop (2000) and
Wandel, Peterson \& Malkan (1999). When necessary, bulge magnitudes
are converted to Johnsons $B$ by adopting $V-R=0.59$ and $B-V=0.93$. The
size of the symbols for the AGNs is proportional to the H$\beta$ FWHM:
the nominal distinction between narrow and broad line AGNs occurs at
the FWHM represented by the symbol size used in the legend. The yellow
line represents the best fit to the nearby quiescent galaxies derived
by Magorrian \etal (1998); the green line is fit to the black hole
masses from Merritt \& Ferrarese (2001a) (shown as filled circles). 
The dotted black lines are the loci for
which the black hole mass is a fixed percentage of the bulge mass. The
arrows in the upper left corner represent the change in $\mh$ or $\mb$
produced under assumptions different from the ones detailed in the
text.}
\end{figure}

We have recomputed the data from the Wandel (1999) and McLure \& Dunlop 
(2000) studies under a uniform set of assumptions, as follows:

\begin{itemize}

\item $H_0=75$ km s$^{-1}$ Mpc$^{-1}$. While black hole masses are
independent of cosmology, bulge magnitudes are not. For nearby
quiescent galaxies with dynamically detected black holes, distances
are estimated directly, mainly using the surface brightness
fluctuation method (SBF).  SBF distances to galaxies in the (mostly)
unperturbed Hubble flow lead to  $H_0=70-77$ km s$^{-1}$ Mpc$^{-1}$
(Ferrarese \etal 2000; Tonry \etal 2001); $H_0=75$ km s$^{-1}$
Mpc$^{-1}$ therefore gives distances, for the distant QSOs and Seyfert
1s, on the same distance scale used for the nearby quiescent galaxies.
Using  $H_0=50$ km s$^{-1}$ Mpc$^{-1}$ instead, as in McLure and
Dunlop, would inflate AGN bulge luminosities by a factor 2.25 (and the
masses by a factor $\sim 2.7$; see below).

\item $R_{BLR}=32.9(\lambda L_{5100}/10^{44} {\rm ergs~s^{-1}})^{0.7}$
light days (Kaspi \etal 2000). While it is likely that the slope of
this correlation will be refined once accurate estimates of 
$R_{BLR}$ are obtained at low and high luminosities, this is currently
the best estimate of the functional form of the $R_{BLR}-L$
relation. Because all QSOs have higher luminosities than the objects
that define the $R_{BLR}-L$ relation, adopting $R_{BLR} \propto
(\lambda L_{5100})^{0.5}$ (e.g. Kaspi \etal 1996) would lead to
estimates of $R_{BLR}$ and $\mh$ that are 1.5 to 3 times smaller
respectively.

\item $v_{\rm BLR}=\sqrt{3}/2{\rm FWHM(H\beta)}$, 
i.e. the BLR is  spherical and characterized by an isotropic velocity 
distribution. This differs from the assumption made by 
McLure \& Dunlop that the BLR is a thin, rotation-dominated disk, 
i.e. $v=1.5{\rm FWHM(H\beta)}$, which predicts velocities $1.7$ times 
larger and black holes masses three times greater.

\item $M/L \propto L^{0.18}$ (Magorrian \etal 1998). This is the
relation defined by the local sample of quiescent galaxies, for which
Merritt \& Ferrarese (2001a) derived $\mh/M_{\rm
bulge}=0.13\%$. Fundamental plane studies (Jorgensen, Franx \&
Kjaergaard 1996) point to a steeper dependence: $M/L \propto
L^{0.34}$.  Accounting for the proper normalization, and given the
range in luminosity spanned by the QSOs and Seyfert 1 galaxies, using
the latter relation  would increase {\it all} inferred $\mh/M_{\rm
bulge}$ ratios by a factor $\sim 2.5$.

\end{itemize}

\begin{figure}
\plotfiddle{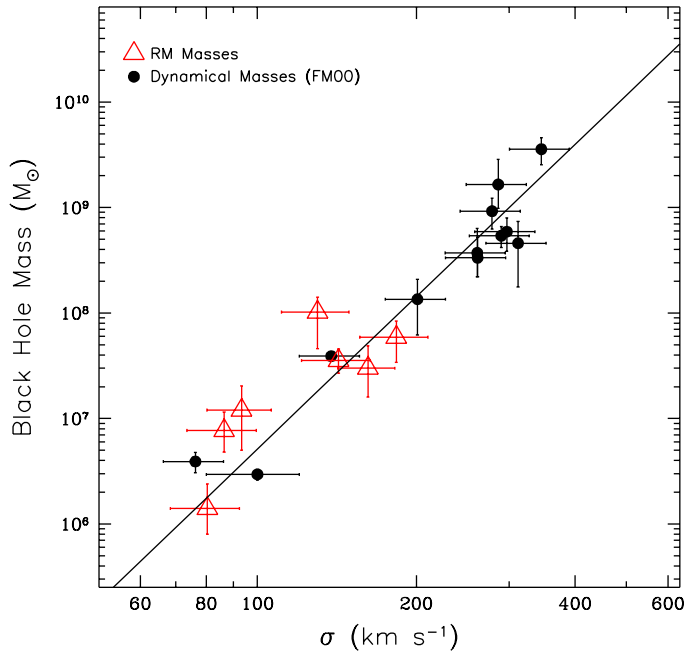}{3.truein}{0.}{50.}{50.}{-160}{-100}
\caption{Black hole mass versus central velocity
dispersion for seven reverberation-mapped AGNs with accurately
measured velocity dispersions, 
compared with the nearby quiescent galaxy sample of Ferrarese \&
Merritt (2000) (plot adapted from Ferrarese \etal 2001).}
\end{figure}

The results are shown in Figure 8.
We draw the following conclusions.

\begin{enumerate}

\item The Seyfert, QSO and quiescent galaxy samples are largely consistent. 
A simple least-squares fit gives $\langle \mh/\mb\rangle=0.09\%$ (QSOs) 
and $0.12\%$ (Seyferts), compared with 
$\langle\mh/\mb\rangle=0.13\%$
for quiescent galaxies (Merritt \& Ferrarese 2001a).
We further note that the disk/bulge decompositions for two of the 
objects with low $\mh/\mb$, $0.001\% - 0.001\%$, 
are deemed of lower quality (McLure \& Dunlop 2000). 
Thus it does not appear to be the case,
as suggested by Richstone \etal (1998) and Ho (1999),
that supermassive black holes in AGN are undermassive 
relative to their counterparts in quiescent galaxies.
In fact, assuming a flattened BLR geometry would further increase 
the AGN masses.

\item $\langle\mh/\mb\rangle$ in AGNs is lower, by a factor
$\sim 6$, than predicted by the Magorrian (1998) relation.
This is further evidence that the mass estimates derived from ground-based
kinematics were systematically in error.

\item In view of recent claims, it is interesting to ask
whether narrow line Seyfert 1s and QSOs (Osterbrock \& Pogge 1985)
contain smaller black holes compared with the rest of the AGN sample
(V\'eron-Cetty, V\'eron \& Goncalves 2001 and references therein;
Mathur \etal 2001). The size of the symbols in Figure 8 is
proportional to the FWHM of the H$\beta$ line: the boundary between
regular and narrow line objects corresponds to the size used in the
figure legend. No correlation between line width and $\mh/\mb$ is
readily apparent for the Seyferts, while a hint might be present for
the QSOs. On the other hand, bulge/disk decompositions are less accurate
for most of the narrow line QSOs, and it is possible that bulge
luminosities in these objects have been overestimated.

\item The large uncertanities in the data, and the large intrinsic
scatter in the $\ml$ relation, make it very difficult to test whether
the relation between $\mh$ and $\mb$ is linear. However, an ordinary least
square fit to the data produces slopes consistent, at the 1$\sigma$
level, with a linear relation for both the QSO and Seyfert 1
samples (cf  Laor 2001).

\end{enumerate}

\subsection{AGN Black Hole Demographics from the $\mh-\sigma$ Relation.}

Because of its large intrinsic scatter, there is little more that can be
learned about black hole demographics from the $\ml$ 
relation.  
An alternative route is suggested by the $\ms$ relation for quiescent
galaxies, which exhibits much less scatter.
Very few accurate measurements of $\sigma$ are available in AGNs, due to
the difficulty of separating the bright nucleus from the faint underlying
stellar population.  The first program to map AGNs onto the $\ms$
relation was undertaken by Ferrarese \etal (2001).  Velocity
dispersions in the bulges of six galaxies with reverberation mapping
masses were obtained, thus producing the first sample
of AGNs for which both the black hole mass and the stellar velocity 
dispersion are accurately known 
(with formal uncertainties of 30\% and 15\% respectively).

Figure 9 shows the relation between black hole mass and bulge
velocity dispersion for the six reverberation-mapped AGNs observed by
Ferrarese \etal (2001), plus an additional object with a high-quality
$\sigma$ from the literature (Nelson \& Whittle 1995). The quiescent
galaxies (Sample A from Ferrarese \& Merritt 2000) are shown by the black dots.
The consistency between black hole masses in active and quiescent
galaxies is even more striking here than in the $\mh-\mb$ plot.
The only noticeable difference between the two samples is a slightly
greater scatter in the reverberation mapping masses (in spite of
similar, formal error bars).
Narrow line Seyfert 1 galaxies do not stand out in any way from the
rest of the AGN sample. 

We conclude that there is no longer any prima facie reason to
believe that reverberation-based masses are less reliable than
those based on the kinematics of stars or gas disks.
This is important since the resolution of stellar kinematical studies
will remain fixed at $\sim 0\Sec1$ for the forseeable future,
whereas reverberation mapping samples a region which is per se
unresolvable and is the only technique that can yield accurate masses
for very small ($\lap 10^6\msun$) or very distant black holes.

\section{Recent and Future Refinements of the $\ms$ Relation} 

\subsection{M33 -- No Supermassive Black Hole?}

The smallest nuclear black holes whose masses have been
securely established are in the Milky Way and M32, both of 
which have $\mh\approx 3\times10^6\msun$ (Table 1).
How small can supermassive black holes be? 
Some formation scenarios (e.g. Haenhelt, Natarajan \& Rees 1998) 
naturally predict a lower limit of $\sim 10^6\msun$. 
Black holes of lower mass have been hypothesized to hide
within off-nuclear ``Ultra Luminous X-Ray Sources'' 
(ULXs: Matsumoto et al.  2001; Fabbiano et al. 2001), 
but their formation mechanisms are envisioned to be completely 
different (e.g. Miller \& Hamilton 2001).
Observationally, a black hole with $\mh \leq 10^6\msun$ could
only be resolved in a very nearby galaxy. 
An obvious candidate is the Local Group late type spiral M33: 
the absence of an obvious bulge, and
the low central stellar velocity dispersion ($\sigma \sim 20$ \kms,
Kormendy \& McLure 1993) both argue for a very small black hole.
According to the $\ms$ relation (Eq. 2), $\mh \sim 3 \times 10^3$
M$_{\odot}$, but a range of at least $1-10 \times 10^3\msun$ is
allowed given the uncertainties in the slope of the relation.

\begin{figure}
\plottwo{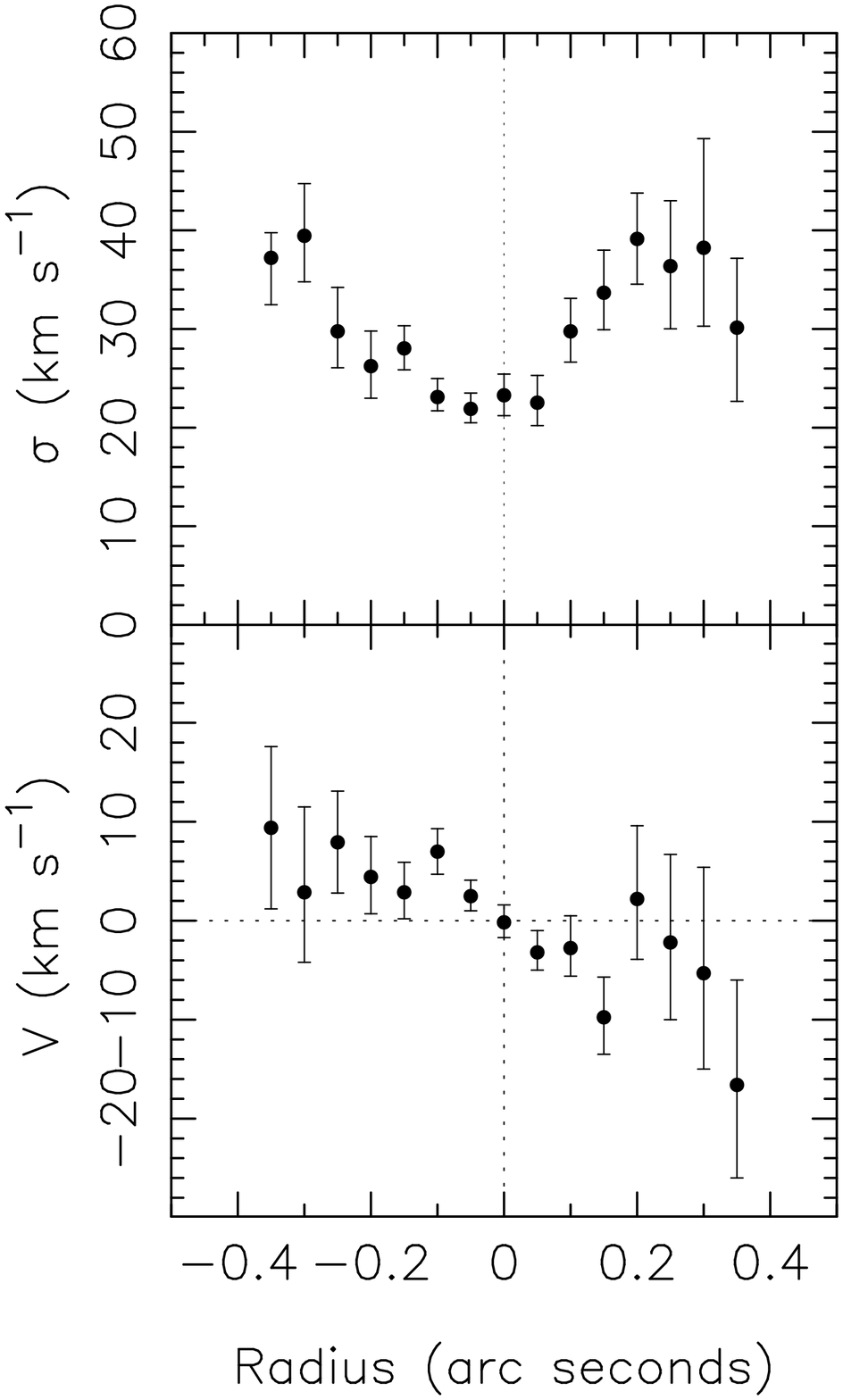}{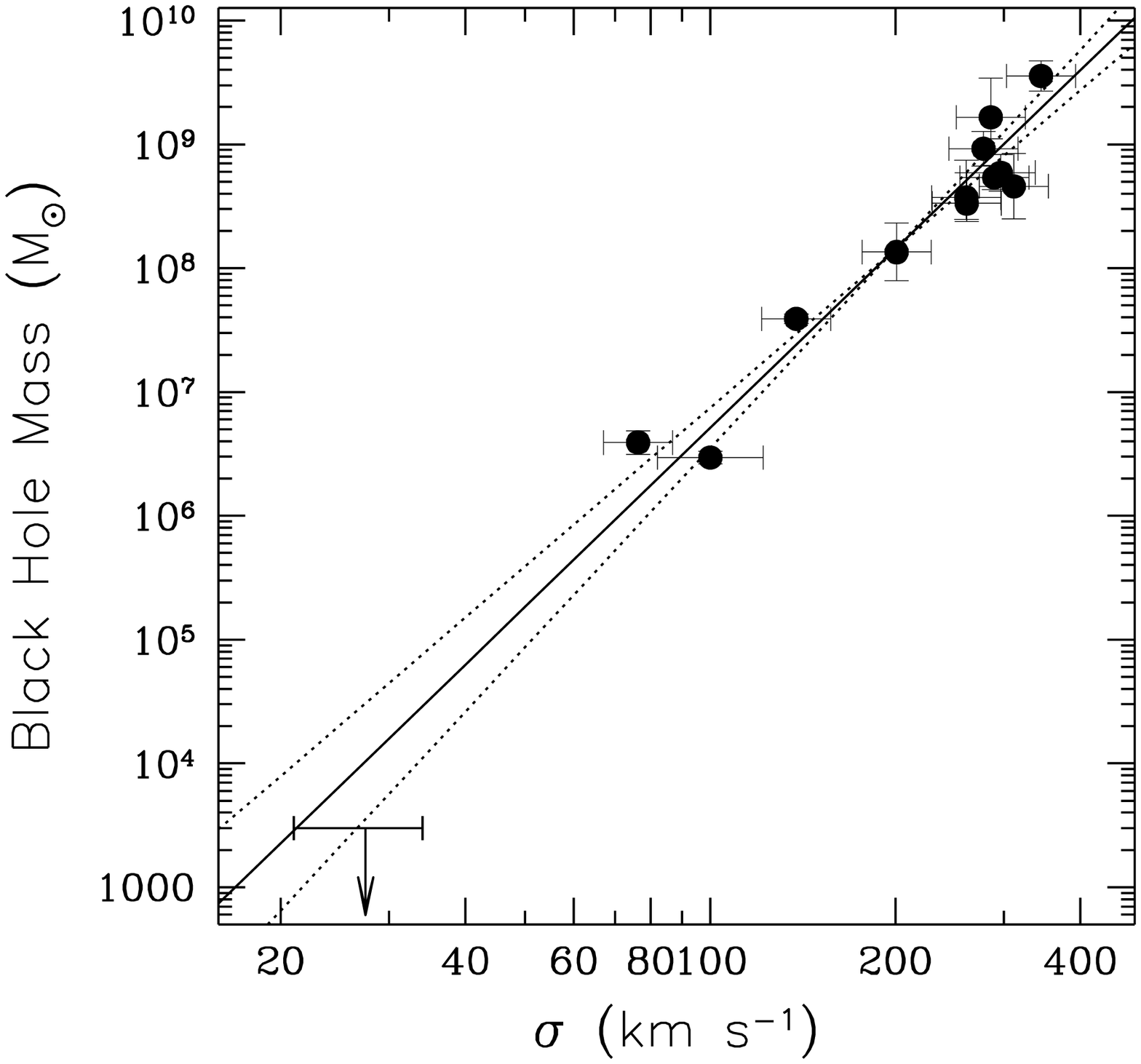}
\caption{
Upper limit on the mass of the black hole in M33
(adapted from Merritt, Ferrarese \& Joseph 2001).
Left: stellar rotation curve and velocity dispersion
profile.
Right: the $\ms$ relation including the upper limit on $\mh$.
}
\end{figure}

We show in Figure 10 the rotation curve and velocity dispersion 
profile of the M33 nucleus obtained from HST/STIS data.
There is an unambiguous {\it decrease} in the stellar velocity
dispersion toward the center of the nucleus: 
the central value is $24\pm 3$ km s$^{-1}$, 
significantly lower than its value of $\sim 35\pm 5$ km s$^{-1}$
at $\pm0.3''\approx 1.2$ pc.
The rotation curve is consistent with solid-body rotation.
A dynamical analysis (Merritt, Ferrarese \& Joseph 2001)
gives an upper limit to the central mass of 
$\sim3 \times 10^3\msun$. 
While this is tantalizingly similar to the masses inferred
for ULXs (Matsumoto et al. 2001; Fabbiano, Zezas \& Murray 2001), the
consistency of this upper limit with the $\ms$ relation (Figure 10 and
Eq. 2) does not allow us to conclude that the presence of a black
hole in M33 would demand a formation mechanism different from the one
responsible for the creation of supermassive black holes in other
galaxies.

Can we expect to hear about the detection of $\leq 10^6\msun$
nuclear black holes in galaxies other than M33 within the next few
years? The remainder of this section summarizes ongoing efforts
and discusses what we are likely to learn.

\subsection{Other New Data from HST}

In the next few years, attempts will be made to detect and
dynamically measure the masses of black holes at the centers
of dozens of galaxies.
The Space Telescope alone is committed to devoting
several hundred orbits to the cause: roughly $130$ galaxies 
have been or will be observed with STIS within the next two years
as part of $\sim 10$ separate projects.

News from some of these projects is already starting to circulate.
Sarzi \etal  (2001) report results from an HST gas dynamical
study of the nuclei of 24 nearby, weakly active galaxies.  Four of the
galaxies were found to have kinematics consistent with the presence of
dust/gas disks (the prototype of which was detected in NGC 4261 by Jaffe
\etal  1994); the authors conclude that in only one of the four galaxies
(NGC 2787) can the kinematics provide meaningful constraints on the presence
of a supermassive black hole. 
Barth \etal  (2001) report the successful detection of
a nuclear black hole in NGC 3245, 
one of six broad-lined AGNs targeted by the team with HST. 
The STIS Instrument Development Team (IDT) has obtained stellar absorption
line spectra for $\sim 12$ galaxies and the data for the first of
these, M32, have been published (Joseph \etal 2001).
The largest sample of stellar dynamical data 
(roughly 40 galaxies, about half of which have already been observed) 
will belong to the ``Nuker'' team.  
Data and a dynamical analysis have been published for one of these 
galaxies (NGC 3379, Gebhardt \etal  2001a) and preliminary masses for
an additional 14 galaxies have been tabulated by Gebhardt \etal (2000b)
and again by Kormendy \& Gebhardt (2001).  Mass estimates were
apparently revised in the second tabulation, some by as much as 50\%.
We adopt the most recent values in the discussion that follows,
pending publication of the full data and analyses.

We can update the $\ms$ relation using the additional black hole
masses that have been published over the last year.
In addition to the 12 galaxies used by Ferrarese \& Merritt (2000) to
define the $\ms$ relation, 10 galaxies listed in Table 1 also have 
${\rm FWHM}/2\rh<1$.
A regression fit accounting for  errors in both coordinates (Akritas \&
Bershady 1996) to the expanded  sample of 22 galaxies gives
\begin{equation}
\mh = (1.48\pm0.24) \times 10^8 \msun \left({\sigma_c\over 200\ {\rm km}\
{\rm s^{-1}}}\right)^{(4.65 \pm 0.48)}
\end{equation}
in good agreement with previous determinations
(Ferrarese \& Merritt 2000; Merritt \& Ferrarese 2001b).
Within this sample, the two subsamples containing only stellar
kinematical or stellar dynamical data produce fits with slopes
of $\sim 4.5$, in agreement with each other and
with the slope quoted for the complete sample.

However, something interesting happens when the eight galaxies in
Table 1 for which $\rh$ has {\it not} been resolved are added to the sample
(all of the mass determinations in these galaxies are based on stellar 
dynamics).
Figure 11 shows the slope of the $\ms$ relation obtained from the stellar
dynamical mass estimates when various cutoffs are placed on the
quality of the data.
If the complete sample is used (including all entries in Table 1
down to NGC 2778), the slope becomes quite shallow, $3.81\pm0.33$.  
When only the best-resolved galaxies are included, ${\rm FWHM}/\rh <
0.2$, the slope increases to $4.48 \pm 0.12$, identical to the value 
obtained from the gas dynamical masses alone.
Mass estimates based on the dynamics of gas disks are 
expected to be more accurate than estimates from stellar dynamics
at equal resolution since the inclination angle of the disk 
can be measured and (if the motions are in equilibrium) the
circular orbital geometry is simpler (Faber 1999).
From Figure 11, we conclude that the inclusion of masses
derived from data that do not properly sample the black hole's sphere of
influence biases the slope of the relation.  A similar conclusion
was reached by Merritt \& Ferrarese (2001b).

We also show in Figure 11 the results of least-squares fits using a
simpler algorithm that does not account for measurement errors.  This
is the same algorithm used by Gebhardt \etal (2000a). 
As pointed out by Merritt \& Ferrarese (2001b), 
not accounting for measurement errors biases the slope too low: 
the inferred slope for the complete sample of stellar dynamical
masses falls to $\sim 3.5$ using the simpler algorithm, similar
to the slope quoted by Gebhardt \etal (2000a) and Kormendy \&
Gebhardt (2001).
Thus the lower slope quoted by those authors is due to the
inclusion of less accurate data points and to the use of a
less precise regression algorithm.

\begin{figure}
\plotfiddle{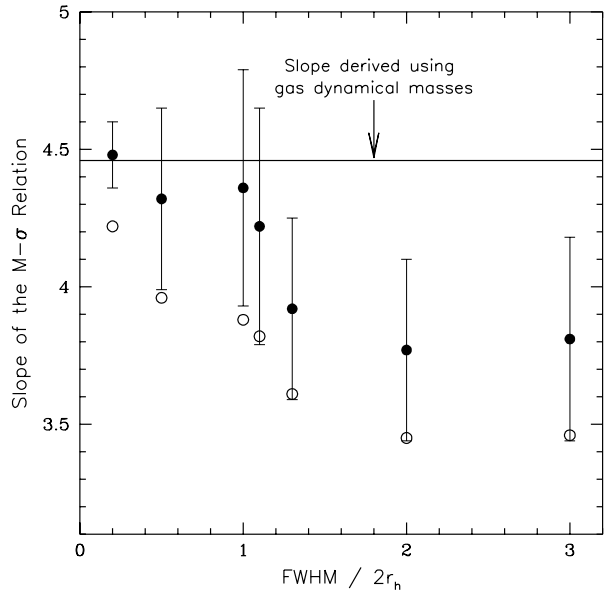}{2.7truein}{0.}{45.}{45.}{-160}{-90}
\caption{The slope of the $\ms$ relation as derived from 
stellar dynamical data as a function of data quality.
Solid circles are slopes derived from a regression algorithm
that accounts for errors in both variables; open circles
are from a standard least-squares routine.
Solid line is the slope derived from gas dynamical data,
for which the black hole's sphere of influence was always resolved.}
\end{figure}

\subsection{The Future}

We noted above that a total of $\sim 130$ galaxies have been
or will be observed with HST during the next two years with the
hope of constraining the mass of the central black hole.
How many of these new data sets will in fact lead to stringent 
constraints on $\mh$?
In Figure 2 we plot the expected radius of influence of the
black hole versus ${\rm FWHM}/2\rh$ for all galaxies for
which a reliable distance (or redshift) and velocity dispersion could
be gathered from the literature. 
Black hole masses have been estimated from
the $\ms$ relation, Eq. (2).  We argued above that
the condition ${\rm FWHM}/2\rh \gap 1.0$, 
compounded by the low  S/N characteristic of HST data,  
will likely lead to weak constraints or biased determinations of $\mh$,
particularly in the case of stellar absorption line data.

Figure 2 shows that the black hole's sphere of influence will
be resolved in less than half of these galaxies. 
Less than one quarter will be resolved as well or better than NGC 3379 
(${\rm FWHM}/2\rh\approx 0.4$), for which the constraints
on $\mh$ are weak (\S2.2).
Among the galaxies slated to be observed in ionized gas,
the preliminary results of Sarzi \etal (2001) suggest that few 
will be found to have the well-ordered disks that are necessary 
for secure estimates of $\mh$.

Many of the ``Sample A'' galaxies from Ferrarese \& Merritt (2000)
were originally targeted for
observation because of their exceptionally favorable properties,
such as nearness (the Milky Way, M32), existence of a
well-ordered maser disk (NGC 4258), etc.
It is unlikely that many more galaxies will turn out 
to have equally favorable properties.

The majority of the targeted galaxies are expected to have black holes 
with masses of order $10^8\msun$. This
range is already well sampled by the current data; however the new
detections might provide useful information about the scatter of the
$\ms$ relation. Only a handful of black holes with masses  
$\gap 10^9\msun$ or $\sim 10^7\msun$ will be detected, 
and probably none in the $<10^7\msun$ range. 
Probing the low and high mass end of the $\ms$
relation is of particular interest since the slope and scatter of the
relation have important implications for hierarchical models of
galaxy formation (Haehnelt, Natarajan, \& Rees 1998; Silk \& Rees
1998; Haehnelt \& Kauffmann 2000) and the effect of mergers on 
subsequent evolution (Cattaneo \etal\ 1999).
The fact that the new observations will not appreciably extend the
range of masses is not due to poor planning: the simple
fact is that very small or very massive black holes are found in
galaxies which are not close enough to resolve their sphere of
influence using current optical/near infrared instrumentation.

It is our opinion that the future of the $\ms$ relation relies on
methods other than traditional dynamical studies. An aggressive
campaign to reverberation map  a large sample of AGNs appears to be
the obvious solution. The recent results from Ferrarese \etal (2001)
show that reverberation mapping  can produce mass estimates with a
precision comparable to traditional dynamical studies. Although the
obvious drawback is that it is only applicable to the $\sim 1\%$ of galaxies
with Type 1 AGN, reverberation mapping is intrinsically unbiased
with respect to black hole mass, provided the galaxies can be monitored with
the appropriate time resolution: while dynamical methods rely on the
ability to spatially resolve the black hole's sphere of influence,
reverberation mapping samples a region which is per se unresolvable.
Furthermore, reverberation mapping can probe galaxies at high redshifts
and with a wide range of nuclear activity, opening an avenue to
explore possible dependences of the $\ms$ relation on redshift and
activity level.

\section{Origin of the $\ms$ Relation}

The greatest dividend to come so far from the $\ms$ relation
has been the resolution of the apparent discrepancy between black hole
masses in nearby galaxies, the masses of black holes in AGN,
and the mass density in black holes needed to explain quasar light.
But the importance of the $\ms$ relation presumably goes beyond 
its ability to clarify the data.
Like other tight, empirical correlations in astronomy,
the $\ms$ relation must be telling us something fundamental about
origins, and in particular, about the connection between black
hole mass and bulge properties.

Probably the simplest way to relate black holes to bulges
is to assume a fixed ratio of $\mh$ to $\mb$.
Since $\mh\propto\sigma^{\alpha}$ (Eq. 2), 
this assumption implies $\mb\propto\sigma^{\alpha}$.
In fact this is well known to be the case: bulge luminosities scale as
$\sim\sigma^4$, the Faber-Jackson law, and mass-to-light
ratios scale as $\sim L^{1/4}$ (Faber \etal 1987), giving
$\mb\sim\sigma^5$, in agreement with the slope $\alpha=4.5\pm0.5$ 
derived above for the $\ms$ relation.

On the other hand, the $\ms$ relation appears to be much tighter than 
the relation between $\sigma$ and bulge mass or luminosity.
And even if a tight correlation between black hole mass and bulge
mass were set up in the early universe, it is hard to see how
it could survive mergers, which readily convert disks to bulges
and may also channel gas into the nucleus, producing 
(presumably) uncorrelated changes in $\mh$ and $\mb$.
The tightness of the $\ms$ relation suggests that some additional 
feedback mechanism acts to more directly connect black hole masses
to stellar velocity dispersions and to maintain that connection
in spite of mergers.

One such feedback mechanism was suggested by Silk \& Rees (1998)
even before the discovery of the $\ms$ relation.
These authors explored a model in which supermassive black holes
first form via collapse of $\sim 10^6\msun$ gas clouds
before most of the bulge mass has turned into stars.
The black holes created in this way would then accrete and radiate,
driving a wind which acts back on the accretion flow.
Ignoring star formation, departures from spherical symmetry etc.,
the flow would stall if the rate of deposition of mechanical energy 
into the infalling gas was large enough to unbind the protogalaxy
in a crossing time $T_D$.
Taking for the energy deposition rate some fraction $f$ of the
Eddington luminosity $L_E$,  
we have
\begin{equation}
f L_E T_D \approx {GM^2_{\rm bulge}\over R_{\rm bulge}}.
\end{equation}
Writing $G\mb\approx\sigma^2R_{\rm bulge}$, $T_D\approx R_{\rm bulge}/\sigma$
and $L_E=4\pi cG\mh/\kappa$ with $\kappa$ the opacity,
\begin{equation}
\mh\approx f^{-1}{\kappa\sigma^5\over 4\pi G^2c}\propto \sigma^5,
\end{equation}
consistent with the observed relation.
The constant of proportionality works out to be roughly correct if 
$f\sim 0.01-0.1$ (Silk \& Rees 1998).

This model assumes that black holes
acquire most of their mass during a fast accretion phase,
$t_{\rm acc}\lap 10^7$ yr.
Kauffmann \& Haehnelt (2000) developed a semi-analytic model 
for galaxy formation in which black holes grow progressively 
larger during galaxy mergers.
The cooling of the gas that falls in during mergers 
is assumed to be partially balanced by energy input from supernovae.
This feedback is stronger for smaller galaxies which has
the effect of steepening the resulting relation between $\mh$
and $\sigma$.
Haehnelt \& Kauffmann (2000) found $\mh\sim\sigma^{3.5}$ but
the slope could easily have been increased if the feedback
had been set higher (M. Haehnelt, private communication).
However the scatter in the $\ms$ relation derived by them was only slightly
less than the scatter in $\mh$ vs $L_{\rm bulge}$, 
in apparent contradiction with the observations (Figure 1).

Burkert \& Silk (2001) also considered a model in which black
holes grow by accreting gas during mergers.
In their model, accretion is halted when star formation begins to 
exhaust the gas in the outer accreting disk; 
the viscous accretion rate is proportional
to $\sigma^3$, and assuming a star formation time scale that is
proportional to $T_D$, Burkert \& Silk found 
$\mh\propto R_{\rm bulge}\sigma^2/G\propto\mb$, with a constant
of proportionality that is again similar to that observed.
This model does not give a convincing explanation for the tight
correlation of $\mh$ with $\sigma$ however.

Feedback of a very different sort was proposed 
by Norman, Sellwood \& Hasan (1996), Merritt \& Quinlan (1998)
and Sellwood \& Moore (1999).
These authors simulated the growth of massive compact objects at
the centers of barred or triaxial systems and noted how the
nonaxisymmetric component was weakened or dissolved when the
central mass exceeded a few percent of the stellar mass.
Since departures from axisymmetry are believed to be crucial for
channeling gas into the nucleus, the growth of the black hole
has the effect of cutting off its own supply of fuel.
These models, being based purely on stellar dynamics, have the 
nice feature that they can be {\it falsified}, and in fact they 
probably have been:
our new understanding of black hole demographics (\S2,3) suggests
that few if any galaxies have $\mh/\mb$ as great as $10^{-2}$.
(At the time of these studies, 
several galaxies were believed from ground-based data
to have $\mh/\mb>1\%$, including 
NGC 1399 (Magorrian \etal 1998), NGC 3115 (Kormendy \etal 1996a), and
NGC 4486b (Kormendy \etal 1997)).

The tightness of the $\ms$ relation must place strong
constraints on the growth of black holes during mergers.
We know empirically that mergers manage to keep galaxies on
the fundamental plane, which is a relation between $\sigma$,
the bulge effective radius $R_e$ and the surface brightness
at $R_e$.
The $\sigma$ that appears in the fundamental plane relation
is the same $\sigma_c$ that appears in the $\ms$ relation (indeed,
it was defined by Ferrarese \& Merritt 2000 for just this reason)
and furthermore $\sigma_c$ is defined within a large enough aperture that
it is unlikely to be significantly affected by dynamical processes
associated with the formation of a black-hole binary during a merger
(Milosavljevic \& Merritt 2001).
Hence the physics of the black-hole binary can be ignored and 
we can ask simply: How do mergers manage
to grow black holes in such a way that $\Delta\log\mh\approx 4.5 \Delta\log\sigma$, independent of changes in $R_e$ and $L$?

This work was supported by NSF grant 00-71099 and
by NASA grants NAG5-6037 and NAG5-9046.
We thank B. Peterson and A. Wandel for useful discussions.

\end{document}